\documentclass[conference]{IEEEtran}
\IEEEoverridecommandlockouts

\usepackage{cite}
\usepackage{amsmath,amssymb,amsfonts}
\usepackage{algorithmic}
\usepackage{graphicx}
\usepackage{textcomp}
\usepackage{xcolor}
\usepackage{algorithm}
\usepackage{array}
\usepackage{ulem} 
\usepackage[caption=false,font=normalsize,labelfont=sf,textfont=sf]{subfig}
\usepackage{textcomp}
\usepackage{stfloats}
\usepackage{url}
\usepackage{verbatim}
\usepackage{epstopdf} 
\usepackage{amssymb}
\usepackage[left=1.62cm,right=1.62cm,top=1.9cm]{geometry}

\def\BibTeX{{\rm B\kern-.05em{\sc i\kern-.025em b}\kern-.08em
    T\kern-.1667em\lower.7ex\hbox{E}\kern-.125emX}}
\begin{document}

\title{RIS-Aided Wireless Communication With Movable Elements: Geometry Impact on Performance\\
\thanks{Y. Zhang is with the Department of Electronic and Electrical Engineering, Trinity College Dublin (e-mail: zhangy42@tcd.ie). This work was supported in part by the Chinese Scholarship Council. This work was also supported in part by Science Foundation Ireland under Grant 13/RC/2077\_P2, and by the EU MSCA Project ``COALESCE'' under Grant Number 101130739.}
}
\author{\IEEEauthorblockN{Yan Zhang}
\IEEEauthorblockA{\textit{dept. Electronic and Electrical Engineering} \\
\textit{Trinity College Dublin}\\
Dublin, Ireland \\
zhangy42@tcd.ie}
\and
\IEEEauthorblockN{ Indrakshi Dey}
\IEEEauthorblockA{\textit{dept. Electronic Engineering } \\
\textit{Trinity College Dublin}\\
Dublin, Ireland \\
Indrakshi.Dey@waltoninstitute.ie}
\and
\IEEEauthorblockN{Nicola Marchetti}
\IEEEauthorblockA{\textit{dept. Electronic and Electrical Engineering} \\
\textit{Trinity College Dublin}\\
Dublin, Ireland \\
nicola.marchetti@tcd.ie}
}

\maketitle

\begin{abstract}
Reconfigurable Intelligent Surfaces (RIS) are known as a promising technology that can improve the performance of wireless communication networks and have been extensively studied. Movable Antennas (MA) are a novel technology that fully exploits the antenna placement to enhance the system performance. 
This article aims to evaluate the impact of transmit power and number of antenna elements on the outage probability performance of an MA-enabled RIS structure (MA-RIS), compared to existing Fixed-Position Antenna RIS (FPA-RIS). The change in geometry caused by the movement of antennas and its implications for the effective number of illuminated elements, are studied for 1D and 2D array structures. Our numerical results confirm the performance advantage provided by MA-RIS, achieving 24\% improvement in outage probability, and 2 dB gain in Signal-to-Noise Ratio (SNR), as compared to FPA-RIS. 
\end{abstract}

\begin{IEEEkeywords}
RIS, Movable antenna, Fixed-position antenna, Performance analysis, Outage probability.
\end{IEEEkeywords}

\section{Introduction}
A comprehensive channel model tailored for Movable Antenna (MA) systems, grounded in field-response principles, is introduced in \cite{zhu2023modeling}. Evaluations of MA system performance enhancements are conducted in \cite{zhu2024performance}, indicating promising outcomes compared to Fixed-Position Antennas (FPA) setups. These enhancements span various metrics such as channel capacity, sum-rate, array gain for beamforming in desired directions, and achievable rate \cite{pi2023multiuser}, achieved through meticulous design and optimization of antenna positions. However, prevailing studies often assume uniform activity across all elements in RIS and mainly consider the MA is deployed at transmitter/receiver/relay, overlooking the intricacies of geometry and element distribution \cite{shao20246dma}.

This simplification overlooks the dynamic nature of RIS operations, where not all elements are active simultaneously. Consequently, it prompts a novel avenue of exploration concerning the accurate identification of active elements, and determining the optimal number needed for peak performance in MA-RIS configurations. This arises from the anticipated deployment of RISs across expansive structures like buildings, where only portions of the RIS area would be illuminated at any given time. In \cite{ntontin2021reconfigurable}, the notion of effective element count is introduced, tied to the illuminated area. The region illuminated by the main lobe of a transmitted beam is typically smaller than the entire MA-RIS area, often approximated as a conical or elliptical shape. Considering that highly directive beams facilitate communication between the Transmitter
(Tx) and MA-RIS, at least 97\% of the transmit energy typically resides within the Half-Power Beamwidth (HPBW) area of the main Tx to MA-RIS beam \cite{ntontin2021reconfigurable}.

The efficiency of RIS is significantly influenced by the geometric arrangement of reflective elements and the compactness of the RIS panel in the near field, as indicated in \cite{ntontin2021reconfigurable}, a notion further supported by \cite{cui2023impact} in the context of 3D RIS deployment. Optimization efforts are carried out in RIS systems equipped with movable antennas \cite{wu2024movable}. However, existing studies primarily focus on FPA-RIS structures, leaving a gap in understanding the impact of changes in the geometry of reflective element deployments in MA-RIS structures.

An inherent challenge in adopting FPA-RIS structures lies in the potential phase distribution offset across various cascaded source-reflective element-destination channels, necessitating distinct patterns of optimal non-uniform discrete phase shifts for each element within the RIS. This phenomenon escalates manufacturing costs, particularly when dealing with a large number of RIS elements \cite{hu2023intelligent}. Conversely, fewer antennas require less physical space for installation, rendering the system more compact and potentially suitable for space-constrained applications. Additionally, the number of elements within an RIS panel significantly influences its output performance. For instance, \cite{zhang2021reconfigurable} examines how the number of elements impacts the system sum rate, aiming to identify optimal configurations. Results indicate a positive correlation between the number of reflecting elements and system sum-rate enhancement.

While larger RIS panels, such as Extremely Large-scale RIS (XL-RIS), offer increased performance potential, they also pose challenges, particularly regarding the passive nature of reflective elements, complicating the acquisition of accurate channel state information \cite{yang2023channel}, a critical requirement for ensuring higher reliability, particularly in Vehicle-to-everything (V2X) systems \cite{xin2022robust}. Training pilot information for XL-RIS can be energy-intensive, necessitating alternative solutions. 

One such potential solution involves integrating MA within the RIS panel, which may achieve comparable or superior system performance while utilizing fewer reflecting elements compared to existing FPA-RIS configurations. This could lead to several advantages: a) \textit{Fewer Reflecting Elements}: Metamaterial antennas can often achieve the same or better performance compared to conventional antennas while using fewer components. This means an MA-integrated RIS panel could potentially use fewer reflecting elements than a standard RIS panel, leading to a more compact and efficient design. b) \textit{Comparable or Superior System Performance}: The unique properties of metamaterials might enable the MA-integrated RIS panel to manipulate radio waves in more sophisticated ways, potentially improving the overall performance of the RIS system in terms of signal strength, beamforming capabilities, or interference reduction.

Driven by the aforementioned considerations, our investigation delves into understanding how the number of elements and transmit power influence outage probability and SNR performance in MA-RIS systems, focusing on sub-6-GHz array antennas in near-field scenarios. The primary contribution of this paper is two-fold. Firstly, we provide a theoretical framework for delineating the illuminated area and determining the effective number of reflecting antenna elements across different MA-RIS configurations (1D versus 2D), factoring in geometric variables such as distance, HPBW, and angle of arrival and departure. Secondly, we integrate environmental blockage effects into our analysis of LoS channels, enabling the derivation of outage probability metrics for varying effective numbers of MA-RIS elements. We conduct a comprehensive comparative assessment elucidating the impacts of MA element count and transmit power on system robustness, with a specific focus on outage probability as a key performance indicator. The results show that the MA-RIS could achieve a lower outage probability than FPA-RIS with a lower number of antennas in both 1D and 2D configurations.

\section{System Model}
\subsection{Channel model}
\begin{figure}[!t]
\centering
\includegraphics[width=1\linewidth]{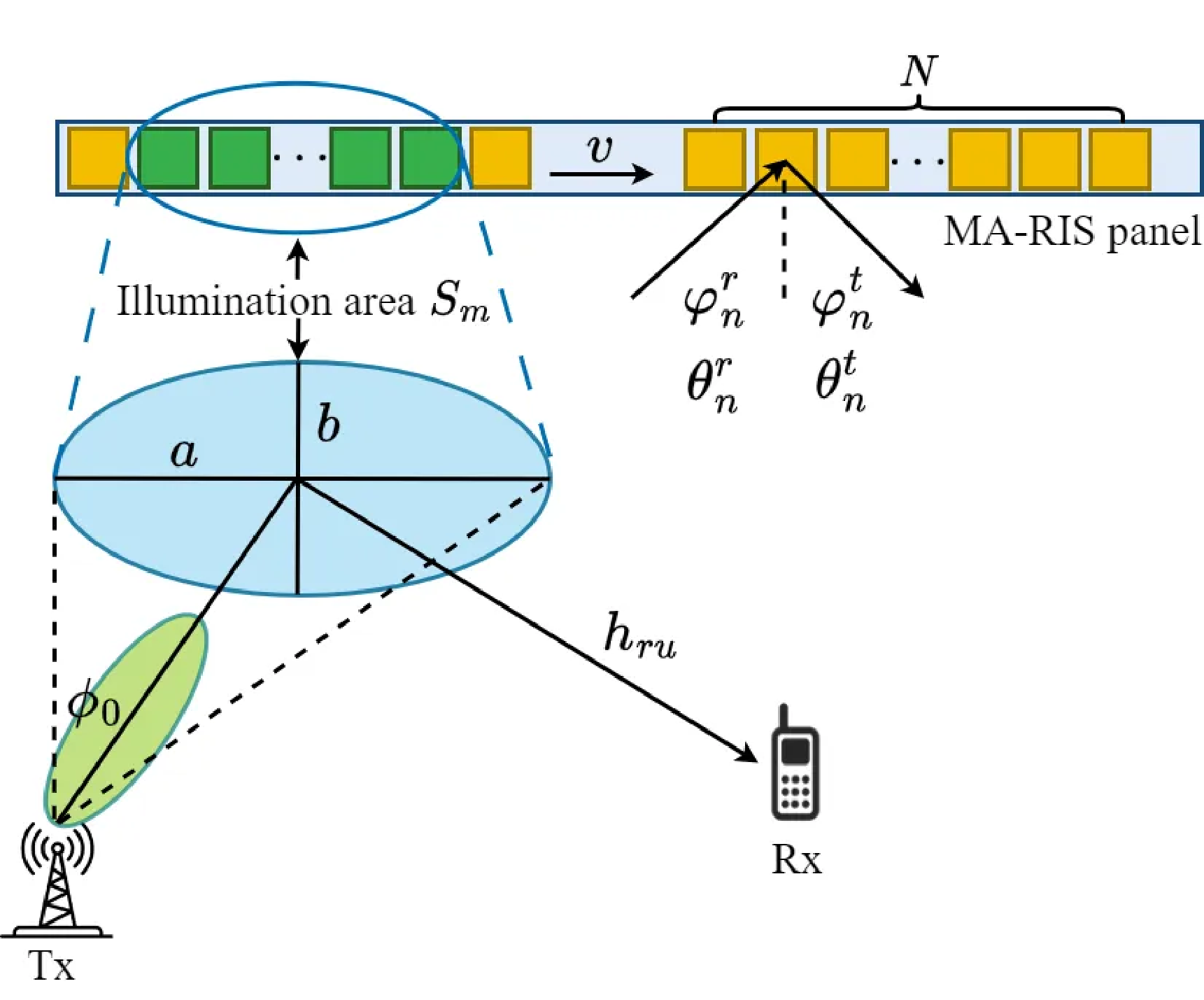}

\caption{Illustration of the system model. In the MA-RIS panel, yellow blocks depict all the $N$ MA units moving with speed $v$, while the green blocks signify the effective antennas. The solid blue ellipse is an enlarged representation of the hollow ellipse which represents the illumination area labeled as $S_m$, with semi-major axis $a$ and semi-minor axis $b$, and represents the illumination area, showing the HPBW $\phi_0$ from the Tx.}
\label{modelill}
\end{figure}

In this section, we introduce the system model and channel model. The system model depicted in Fig.~\ref{modelill}, we model the partial RIS illumination as an elliptical area, which is essential for determining the effective number of elements \cite{stratidakis2022understanding}. 

The MA-RIS panel, featuring with a certain length, enables 1D and 2D alignment MAs moving horizontally from the left panel edge to the right panel edge. The 1D and 2D structures are equipped with $1 \times N$ and $N_x \times N_y$ MAs separately, where we let $N = N_x \times N_y$ to make sure the 1D and 2D structured MA-RIS have the same number of MAs. 

We consider the simplified LoS scenario with only one transmit and receive path between Tx and RIS, and the same for the RIS to Rx path. The position of Tx can be represented as $\textbf{b} = [x_{b}, y_{b},  0]^T$. Thus, the normalized wave vector of one single transmit path can be represented by $\textbf{n} = [\mathrm{sin} \theta_b \mathrm{cos}\varphi_{b}, \mathrm{cos} \theta_b, \mathrm{sin} \theta_b \mathrm{sin}\varphi_{b}]^T$, where $\theta_b$ and $\varphi_{b}$ represent the Elevation of Departure (EoD) and Azimuth of Departure (AoD) from the Tx to MA-RIS link, respectively. Therefore, following \cite{qin2024antenna}, the relative path difference from $\textbf{b}$ to reference point $\textbf{b}_{0} = [0, 0, 0]^T$ is denoted as $\rho(\textbf{b}) = \textbf{n}^T(\textbf{b}-\textbf{b}_{0}) 
= x_{b}\mathrm{sin} \theta_b \mathrm{cos}\varphi_{b} + y_{b}\mathrm{cos} \theta_b $. The resulting phase difference is $\frac{2 \pi}{\lambda }\rho(\textbf{b})$. Moreover, the array steering vector at Tx is defined as follows
\begin{equation}
    s(\textbf{b}) =e^{j\frac{2 \pi}{\lambda } \rho(\textbf{b})}= e^{j\frac{2 \pi}{\lambda }\left ( x_b\mathrm{sin} \theta_b\mathrm{cos}\varphi_b + y_u\mathrm{cos} \theta_b\right) }\,,
\end{equation}
where $\lambda$ is the carrier wavelength.

The center position of the MA-RIS panel is denoted as $\textbf{r}_c = \left[x_r, y_r, z_r \right]^T$. The position of the $n$-th MA is represented as $\textbf{r}_n = \left[x^r_{n},  y^r_{n}, z^r_{n}\right]^T\in C_r$ for $n \in \mathcal{N} \triangleq \left \{ 1,\dots ,N\right\}$, where $C_r$ denotes the region within which the MAs can move freely \cite{cheng2023movable, zhu2023modeling}.  Therefore, we get the path difference from Tx to the $n$-th MA as $g^r(\textbf{r}_n) = e^{j\frac{2 \pi}{\lambda } \rho ^r(\textbf{r}_{n})}=e^{j\frac{2 \pi}{\lambda } \left ( x^r_{n}\mathrm{sin} \theta_{n}^{r}\mathrm{cos}\varphi_{n}^{r} + y^r_{n}\mathrm{cos} \theta_{n}^{r}\right)}$, where $\theta^r_n$ and $\varphi_{n}^{r}$ respectively represents the Elevation of Arrival (EoA) and Azimuth of Arrival (AoA) for each MA-RIS element, and $\rho^r (\textbf{r}_{n})$ is the path difference from $\textbf{r}_{n}$ to reference point relative to the received path. 
By stacking together the $g^r(\textbf{r}_n)$ of all $N$ MAs, we obtain the collective array response vector of the receive MA is defined as follows
\begin{equation}
    \textbf{G}^r(\textbf{r}) \triangleq \left [g^r(\textbf{r}_1), g^r(\textbf{r}_2),\dots ,g^r(\textbf{r}_N)  \right ] \,. 
\end{equation}
Similarly, by defining $g^t(\textbf{r}_n) =e^{j\frac{2 \pi}{\lambda } \rho^t (\textbf{r}_{n})}= e^{j\frac{2 \pi}{\lambda } \left ( x^r_{n}\mathrm{sin} \theta_{n}^{t}\mathrm{cos}\varphi_{n}^{t} + y^r_{n}\mathrm{cos} \theta_{n}^{t}\right) }$, and $\rho^t (\textbf{r}_{n})$ is the path difference from $\textbf{r}_{n}$ to reference point relative to the transmitted path. The array response of MA-RIS with $\theta^t_n$ and $\varphi_{n}^{t}$ representing the EoD and AoD from $n$-th MA-RIS element to Rx link is defined as
\begin{equation}
    \textbf{G}^t(\textbf{r}) \triangleq \left [g^t(\textbf{r}_1), g^t(\textbf{r}_2),\dots ,g^t(\textbf{r}_N)  \right ] \,.
\end{equation}
The Rx is fixed at $\textbf{u} = \left[x_{u}, y_{u}, z_{u}\right]^T$ and $\theta^u$ and $\varphi^u$ represent the EoA and AoA to Rx. Then with the path difference $\rho^t (\textbf{u})$, the corresponding FRV relative to the reference point is expressed as
\begin{equation}
    f(\textbf{u}) =e^{j\frac{2 \pi}{\lambda } \rho^t (\textbf{u})}= e^{j\frac{2 \pi}{\lambda }\left ( x_u\mathrm{sin} \theta^u\mathrm{cos}\varphi^u + y_u\mathrm{cos} \theta^u\right) }\,.
\end{equation}
The small-scale fading channel vector for the Tx to MA-RIS link can then be written as
\begin{equation}
    \rm {\mathbf{h}}_{br} = \alpha_n \mathbf{G}^r(\mathbf{r})^H s(\mathbf{b}) \,,
\end{equation}
while the channel vector for the MA-RIS to Rx link can be written as
\begin{equation}
    \rm {\mathbf{h}}_{ru} = \beta_n f(\mathbf{u}) \mathbf{G}^t(\mathbf{r})\,,
\end{equation}
where $\alpha_n$ and $\beta_n$ represent the channel's response coefficients. 

Denoted by $ {\rm s} \in \mathbb{C}^{ M \times 1} $ the input of the  downlink transmission, then the received signal at Rx is expressed as
\begin{equation}
    y_k =  \sqrt{P_r} {\rm{ {\mathbf{h}}_{ru}\boldsymbol{\Theta} \rm {\mathbf{h}}_{br} s}} + n_0 \,,
\end{equation}
where $P_r$ is the received power,  $n_0$ is the additive complex Gaussian noise with zero mean and variance of $\sigma_n^2$, and $\boldsymbol{\Theta} = \mathrm{diag} \left\{e^{j \theta_1}, \cdots,e^{j \theta_N} \right\}$ is the phase shift matrix of MA-RIS.  Letting $\varpi=\rho(\textbf{b})+\rho^r(\textbf{r}_{n})+\rho^t(\textbf{r}_{n})+\rho^t(\textbf{u})$, following the SNR maximization rule in \cite{cheng2023sum},  $y_k$ can be expanded as
\begin{align}    \label{eq}
y_k &= \sqrt{P_r} { \rm{ {\mathbf{h}}_{ru}\boldsymbol{\Theta} \rm {\mathbf{h}}_{br} s}} + n_0 \nonumber \\
&= {\sqrt{P_r }} \sum_{n=1}^{N_{eff}} \alpha_n \beta_n e^{j \left ( \theta_i-\frac{2 \pi}{\lambda } \varpi \right) }  s + n_0 \,, \nonumber  \\
\end{align}
where $N_{eff}$ is the effective number of elements determined by the illumination area in MA-RIS, leading to the Rx's received instantaneous SNR as
\begin{equation} \label{eq2i}
    \gamma = {\frac{P_r}{ \sigma_n^2}}\left | \sum_{n=1}^{N_{eff}}\alpha_n \beta_n e^{j \left ( \theta_i-\frac{2 \pi}{\lambda }\varpi  \right )  } \right |^2 \,.
\end{equation}
By inspecting \eqref{eq2i}, we note that $\gamma$ reaches its maximum value in correspondence of
\begin{equation} \label{eq2}
    \theta_i=\frac{2 \pi}{\lambda }\left ( \rho(\textbf{b})+\rho^r(\textbf{r}_{n})+\rho^t(\textbf{r}_{n})+\rho^t(\textbf{u}) \right)\,.
\end{equation}
Hence, we can get the maximum SNR as $\gamma_{max} = A^2 \bar{\gamma}$, where $A = \sum_{n=1}^{N_{eff}}\alpha_n \beta_n $, and the average SNR is $\bar{\gamma}={\frac{P_r}{ \sigma_n^2}}$.

\subsection{Topology of MA-RIS illumination area}
Let us consider the system model shown in Fig.~\ref{modelill}. All the elements are moving horizontally, with the same constant speed $v$ from the RIS's left panel edge toward the right edge. By fitting the illumination area into an elliptic shape, we can derive the illumination area that affects the Rx's SNR.
In order to capture the most important feature in the context of MA-RIS  -- the angle of arrival and departure information related to the movement of MA -- we followed the cosine model in \cite{yu2017coverage, cui2023impact}, and for the $n$-th element of MA-RIS, its radiation pattern for the incident beam from Tx is then given by $U_t(\varphi_{n}^{r}) = \mathrm{cos}^2\left(\frac{\pi (\varphi_{n}^{r}-\varphi_{r})}{2\phi_0 } \right)$, where the HPBW of the transmitted beam is denoted as $\phi_0$, with the angular information $\varphi_{n}^{r} = \mathrm{arctan}\left(x_n^r/y_n^r\right)$, and with $\varphi_r = \mathrm{arctan}\frac{x_r }{y_r}$ representing the AoA for MA-RIS center. The radiation pattern of the $n$-th element of MA-RIS corresponding to the incident and reflected beam, is separately defined in a similar way as in \cite{yu2017coverage}, as $W_r(\varphi_{n}^{r}) = \mathrm{cos}^3(\varphi_{n}^{r})$ and $W_t(\varphi_{n}^{t}) = \mathrm{cos}^3(\varphi_{n}^{t})$ with $\varphi_{n}^{t} = \mathrm{arctan}\frac{x_{u} - x^r_{n} }{y_{u} - y^r_{n}}$ representing the AoD from $n$-th MA-RIS element to Rx. We deploy the near field path loss model in \cite{tang2020wireless} and get the received power at the Rx as
\begin{equation}
    P_r = P_t \frac{\lambda^2 d_x d_y}{64 \pi^3}\left | \sum_{n=1}^{N_{eff}} \frac{\sqrt{U_r(\varphi_{n}^{r})W_r(\varphi_{n}^{r})W_t(\varphi_{n}^{t})} }{d_{n,1} d_{n,2}}  \right |^2 \,,
\end{equation}
where $d_x \times d_y$ represents the size of a flat MA-RIS element, $d_{n,1}= \left \| \textbf{b} -\textbf{r}_n \right \|$ and $d_{n,2}=\left \| \textbf{r}_n - \textbf{u} \right \|$ denote the 3D distance between Tx to $n$-th MA-RIS element and the 3D distance between the $n$-th MA-RIS element to Rx, respectively.

We model the illumination area of the MA-RIS as an elliptical shape with area size as $S_m = \pi a b$. The semi-major and semi-minor axes of the ellipse area are calculated as $a = \frac{d_c\mathrm{sin} \phi_0 }{\mathrm{sin} (\varphi_t + \phi_0 )}  $ and $b = \frac{d_c \mathrm{sin} \phi_0 }{\mathrm{cos} (\theta_r + \phi_0 )}$ where $d_c$ is the 3D distance between Tx and the center of the MA-RIS panel, and $\theta_r$ is the EoA from Tx to MA-RIS's center. Therefore, we can get the effective illuminated number of the 1D and 2D MA-RIS as $N^{1D}_{eff} = \left \lceil \frac{a}{d_s}  \right \rceil$, $N^{2D}_{eff} = \left \lceil \frac{S_m}{2d_s^2}  \right \rceil$, where $d_s$ is the inter-element spacing, and $\left \lceil \cdot  \right \rceil$ represents the ceiling function.
\subsection{Outage Probability}
Outage probability is a widely used metric for assessing performance in communication systems \cite{cui2023impact}. It is deﬁned as the probability that the SNR is smaller than a given threshold $\gamma_{th}$, 
\begin{equation}
    P_{out}(\gamma_{th}) = \mathrm{Prob}(\gamma_{max}\le \gamma_{th}) = F_{\gamma_{max}}(\gamma_{th}) \label{p_out} \,,
\end{equation}
where $F_{\gamma_{max}}(\cdot)$ denotes the Cumulative Distribution Function (CDF) of $\gamma_{max}$, where $\gamma_{max} = \bar{\gamma} \left (  {\textstyle \sum_{n=1}^{N_{eff}} \eta_n}  \right )^2$, where we denote $\eta_n = \alpha_n\beta_n$. For simplicity, we adopt the description same as \cite{cui2021snr}, which allows us to obtain $F_{\gamma_{max}}=\frac{\Gamma \left ( \Lambda N_{eff},\sqrt{\frac{\gamma_{th}}{\bigtriangleup^2 \bar{\gamma}} }  \right ) }{\Gamma\left ( \Lambda N_{eff} \right ) }$, where $\Gamma \left ( \cdot,\cdot \right )$ is the incomplete Gamma function, while $\Gamma \left ( \cdot \right )$ is the Gamma function, parameters are set as $ \Lambda =\frac{\pi^2}{16-\pi^2}$, and $\bigtriangleup = \frac{\left(16-\pi ^2\right) }{2 \pi}$ following \cite{cui2023impact}. 
\begin{figure}[H]
\centering
\subfloat[]{\includegraphics[width=3.5in]{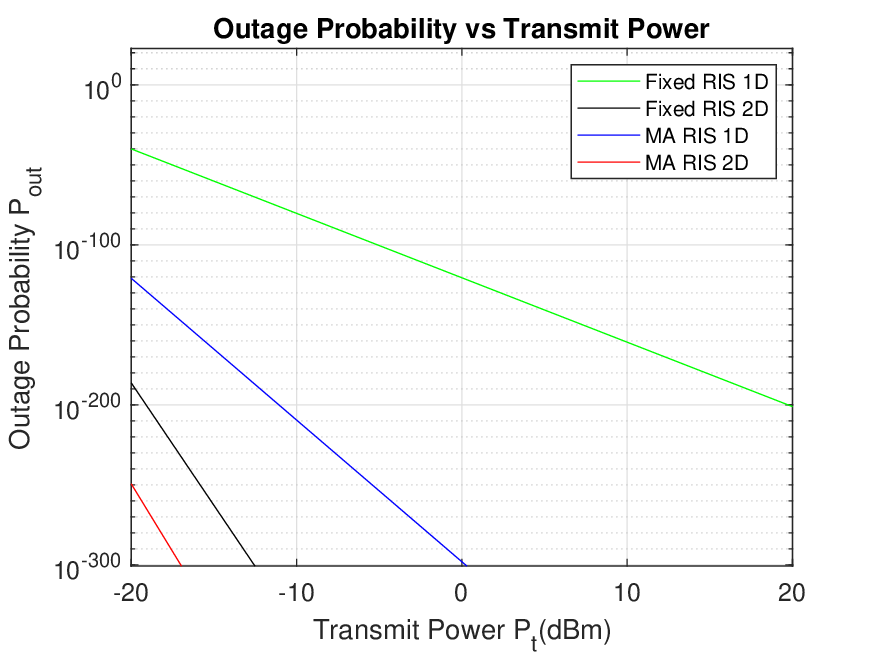}
\label{OutvsPt}}
\hfil

\subfloat[]{\includegraphics[width=3.5in]{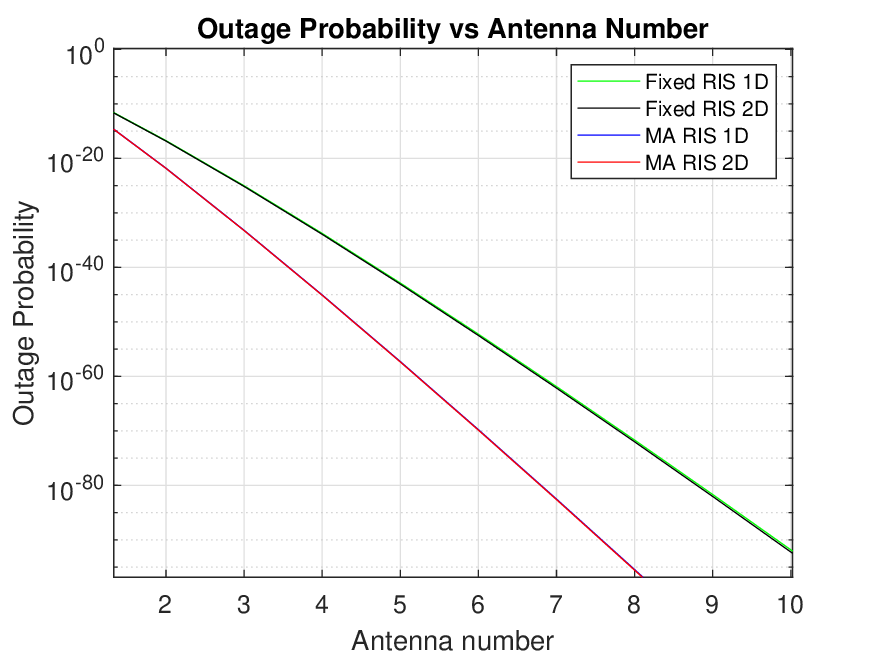}%
\label{OutvsN}}
\caption{Outage probability $P_{out}$ performance comparison where HPBW $\phi_0=13.8^{\circ} $, $y_s= 15$m, and $\gamma_{th}=20$ dB: (a) $P_{out}$  versus transmit power $P_t$, (b) $P_{out}$  versus elements number $N$, where transmit power $P_t= -20 $ dBm.}
\label{OutageProb}
\end{figure}
\section{Performance Analysis}
We analyze the performance in the near field and  assume that the elements within the MA-RIS configuration shift horizontally, transitioning from the panel's left edge to its right edge. The positions of Tx and Rx are set as $(x_b, y_b, z_b) = (0, 0, 3) $ m and $(x_{u}, y_{u}, z_{u}) = (10, 0, 0) $ m, The RIS panel is on the xy-plane, so $z_c =0$ m, x-axis's center position is  $x_c = 5$ m, $y_c $ changes along with the RIS panel's position during each iteration in y-axis, the length of RIS panel is set as $l=1$m, and the number of elements is $N = 21$. The antenna movement velocity is $v=0.1$ m/s, with the traversal time thus amounting to $T=10$s. Following \cite{tang2020wireless}, we set the carrier frequency as $f = 4.25$ GHz, and the noise variance as $\sigma^2 = -45$ dBm. 
\begin{figure}[H]
    \centering
    \includegraphics[width=1\linewidth]{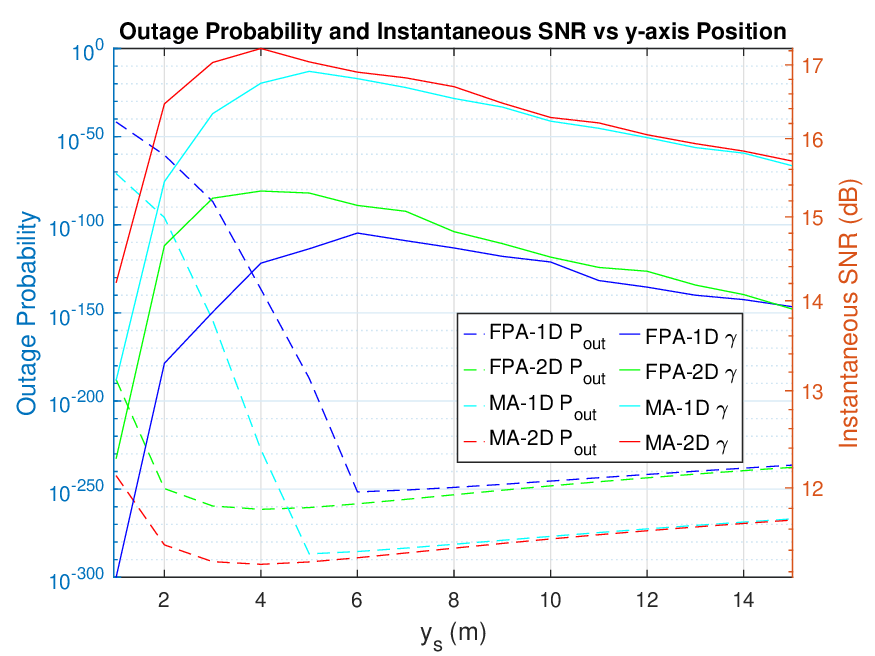}
    \caption{Outage probability $P_{out}$ and Instantaneous SNR $\gamma$ versus y-axis position $y_s$ for HPBW $\phi_0=13.8^{\circ} $, and $\gamma_{th} = 20$ dB. }
    \label{OutInsvYs}
\end{figure}
We assume that all the $N $ MA elements move together with speed $v$ along the RIS panel of length $l$ m within a duration time of $T$ s. The RIS panel is symmetrically set at $\textbf{r}_c$ on the xy-plane while it also moves vertically along y-axis. So we can write the offset from the RIS panel's center $x_c$ at each time instant as $d_{t,\mathrm{off}}= (t-1)v - l/2$, where $t \in \left \{1,2, \cdots, T\right \}$. So we could get the center element's position at each time instant $x_{t,c}  = x_c + d_{i,\mathrm{off}}$, and in turn each MA element's x-axis position $x^r_n = x_{t,c} + (n-1)*d_{sm}, n \in \left \{1,2, \cdots, N\right \}$, where $d_{sm}$ is the element spacing between MAs.

\renewcommand\thesubsection{\Alph{subsection}}
\subsection{\textit{Transmit Power and Number of Elements : Effect on Outage Probability}}
As shown in Fig.~\ref{OutageProb}\subref{OutvsPt},  the outage probability decreases with an increase in transmit power for all configurations. In 2D scenario with 21 elements and -20 dBm transmit power, the MA-RIS achieves around 24\% better performance than FPA-RIS in terms of outage probability. In a 2D scenario, 

the outage probability sharply decreases with  increasing transmit power due to better control over signal propagation and interference mitigation. This is because the additional spatial dimension enables more precise optimization of signal paths, resulting in improved channel conditions. The impact of number of elements on outage probability $P_{out}$ is shown in Fig.~\ref{OutageProb}\subref{OutvsN}, with the transmit power $P_t$ fixed at -20 dBm. Due to the single-path scenario we consider, there is no significant difference between the 1D and 2D configurations for both the MA-RIS and FPA-RIS settings, due to the fact the outage probability is jointly determined by $\bar{\gamma}$ and $N_{eff}$ as $P_{out}\to \left ( \frac{\bar{\gamma}\triangle }{\gamma_{th}}\Gamma \left ( \Lambda N_{eff}\right )^{-\frac{2}{\Lambda N_{eff}} }  \right )^{-\frac{\Lambda N_{eff}}{2} } $ \cite{cui2023impact}. As we can see from Fig.~\ref{OutInsvYs}, when the MA-RIS panel reaches the highest heights, $P_{out}$ and $\gamma$ gradually approach a common value for the 1D and 2D cases, matching what is shown in Fig.~\ref{OutageProb}\subref{OutvsN}.  In particular, the results show that MA-RIS uses fewer elements than FPA-RIS to achieve the same level of performance, for example for an outage probability of $10^{-70}$, MA-RIS uses 25\% fewer elements than FPA-RIS. 

The maximum instantaneous SNR $\gamma$, was determined at positions $y_s = 4.5, 4, 6, 4$ m, yielding instantaneous SNR values of 16.8, 17.2, 14.8, and 15.3 dB for MA-1D, MA-2D, FPA-1D, and FPA-2D, respectively. 

These variations in the optimal positions at $y_s$ are attributed to differences in geometric compactness. Given that channel state information is assumed to be known, the average SNR, $\bar{\gamma}$, follows a similar trend to the instantaneous SNR, $\gamma$. Consequently, the $P_{out}$ is inversely proportional to the instantaneous SNR and is influenced by both $\gamma$ and the effective number of elements $N_{eff}$ based on \eqref{p_out}. As $N_{eff}$ increases within a relatively lower height range, $P_{out}$ initially decreases; however, once $N_{eff}$ reaches its maximum, $P_{out}$ begins to increase as \(\gamma\) decreases.

\par These promising results are the consequence of MA-enabled dynamic reconfiguration of antenna elements, allowing for adaptive adjustment of the antenna pattern and phase distribution. This flexibility can be leveraged to optimize the signal path and mitigate multipath fading, leading to enhanced communication reliability and lower outage probabilities compared to FPA-RIS, which instead have a limited capability to adapt to changing channel conditions. 

\subsection{\textit{Effective Number of MA-RIS}}
\begin{figure}[htbp]
    \centering
    \includegraphics[width=1\linewidth]{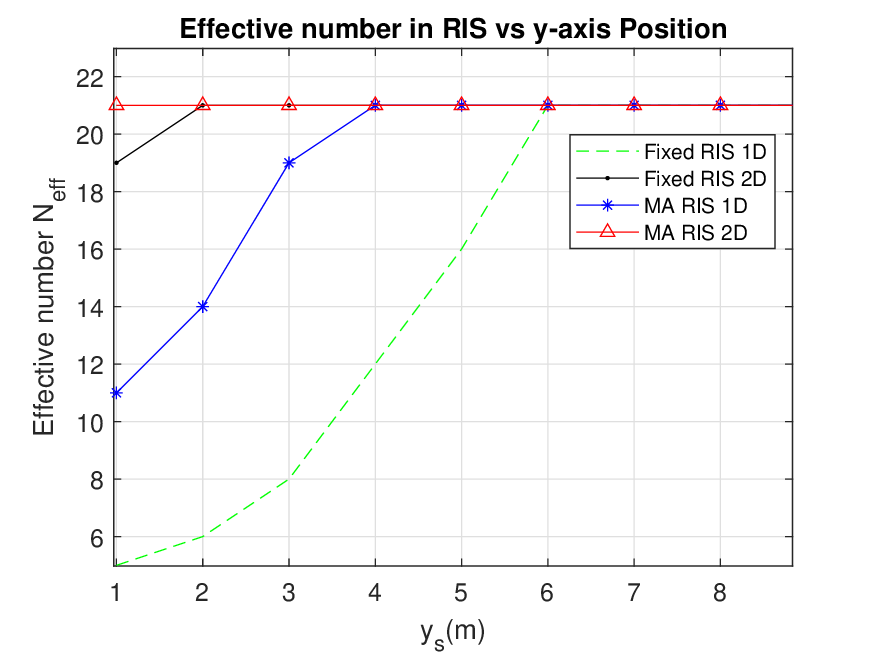}
    \caption{Comparisons of effective number of elements in MA-RIS and FPA-RIS versus y-axis position $y_s$ where HPBW $\phi_0=13.8^{\circ} $, the center position of RIS panel is $x_c= 5$ m along x-axis.}
    \label{fig:Neff}
\end{figure}

We assume that both the MA-RIS and FPA-RIS panels move vertically along the y-axis, denoted as $y_s=y_r$ in Fig.~\ref{fig:Neff}. They move from an initial position of $y_s=$1 m to the maximum height of $y_s=15$ m in increments of 2 m in xy-plane. In order to compare the performance of FPA-RIS and MA-RIS over the same overall panel length, the antenna spacing for FPA-RIS is set as $d_{sf} = \lambda/2$, while the antenna spacing for MA-RIS is $(vt-(N-1)d_{sf})/(N-1)$. 

We can observe that for both 1D and 2D configurations for the MA-RIS, the effective antenna number $N_{eff}$ reaches its maximum value for lower values of $y_s$ as compared to FPA-RIS. This indicates that the ability to adjust the position of the RIS panel allows one to optimize the reflective path and thus the effective utilization of more antenna elements. We also notice that in 2D configurations, the RIS can quickly reach its maximum $\textit{N}_{eff}$ as it begins to move from the initial position along the y-axis. This implies that, the 2D RIS has a higher initial capacity for effective signal reflection due to its expanded interaction field, reaching peak performance with minimal movement. Instead, for 1D configurations, the effectiveness of the RIS in terms of $\textit{N}_{eff}$ increases more significantly as one moves to higher positions along the y-axis. In fact, in 1D setups the initial positions may not be optimal for capturing and redirecting signals effectively, while as the RIS moves to higher y-axis positions, it can encounter angles and positions where the linear array of antennas becomes more aligned with the direction of the incident signals, thereby increasing the effective number of antennas that can contribute to signal reflection. 

\begin{figure}[htbp]
    \centering
    \includegraphics[width=1\linewidth]{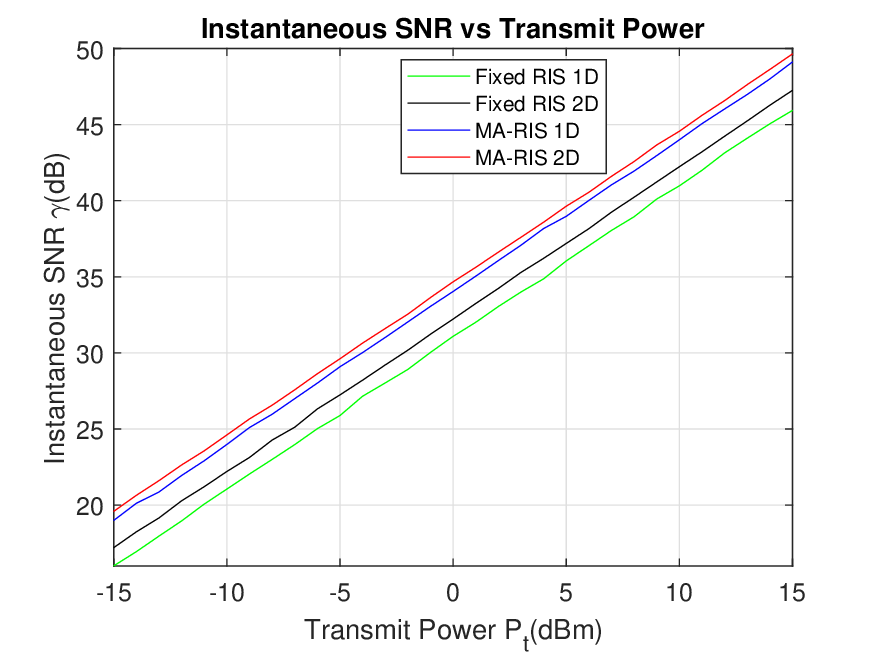}
    \caption{Instantaneous SNR $\gamma$ versus transmit power $P_t$, with noise variance $\sigma^2 = -45$ dBm, and y-axis position $y_s=15$ m.}
    \label{fig:InsSNR}
\end{figure}
\subsection{\textit{Instantaneous SNR}}
In Fig.~\ref{fig:InsSNR}, we compare instantaneous SNR versus the transmit power $P_t$. We can see for example that when the transmit power $P_t$ is 0 dBm, then the MA-RIS's instantaneous SNR is about 3 dB and 2.5 dB higher than FPA-RIS in 1D and 2D setups, respectively. This is due to the fact that MA-RIS systems have the inherent advantage of being able to adapt their position and orientation, in response to the environment and the direction of the incoming signal. This adaptability allows them to optimize the reflection path, enhancing the signal's strength at the receiver and thereby improving the SNR. The superior performance of 2D is leaded by the geometry compactness. 2D configurations can enable more sophisticated beamforming strategies, leveraging the additional spatial dimension to focus the energy more precisely towards the receiver. Moreover, 2D RIS configurations can accommodate a wider range of AoA and AoD values for the signal. This capability enables them to maintain high SNR levels over a variety of transmission and reception scenarios, unlike 1D RIS, which is more limited in this respect. 

\section{Conclusion}
In this article, we derived the effective illumination area of MA-RIS systems. Our results show that the MA-RIS configuration generally provides a performance advantage over FPA-RIS, particularly in 2D setups. The consistently higher effective number of elements in  2D MA-RIS, the higher SNR, and the lower outage probability, all highlight the benefits of incorporating reflective element mobility into RIS design. 
\par Our work highlights the importance of considering the following factors with regard to antenna elements: the element geometry distribution, i.e., 1D or 2D, and each element's position in RIS panel, as well as the position of the RIS panel. Those factors are deeply connected with the angles of incidence and interaction with reflection signals, which contribute the signal strength and coverage. 
This could prove particularly beneficial in dynamic or large-scale environments, e.g., V2X and Massive MIMO communication systems, since the flexible adjustment of the antenna positions could provide an important additional degree of freedom to capture the change in environment characteristics, resulting in enhanced coverage, and improved signal quality. Despite our work highlight the advantages of MA-RIS, the limitations inherent in our current model set the stage for our future efforts to refine and expand its capabilities. The model's idealized assumptions, such as a fixed single-user position and a single-path scenario, led to a notably low outage probability. Our future work aims to address these limitations by incorporating more realistic multi-user mobility scenarios and multipath propagation conditions, enhancing both the model's applicability and accuracy. 


\normalem
\bibliographystyle{unsrt}
\bibliography{Citation1}

\end{document}